\documentclass[]{aa}
\usepackage{times}
\usepackage[normalem]{ulem}
\usepackage{mathptm}
\usepackage{amssymb}
\usepackage[dvips]{epsfig}
\usepackage[dvips]{rotating}
\usepackage{natbib}
\usepackage{times}
\bibpunct{(}{)}{;}{a}{}{,}

\newcommand{\tefft}{$T_{\mbox{\scriptsize eff}}$}
\newcommand{\teffm}{T_{\mbox{\scriptsize eff}}}

\begin{document}     

\title{The stellar content of the Hamburg/ESO survey\thanks{Based on
    observations collected at the European Southern Observatory, La
    Silla (Proposal No. 66.B-0054), and Kitt Peak National and
    Cerro Tololo Inter-American Observatories, operated by the Association
    of Universities for Research in Astronomy, Inc., under cooperative
    agreement with the U.S. National Science Foundation.}}
    
\subtitle{III. Field horizontal-branch stars in the Galaxy}

\author {N. Christlieb\inst{1}
\and T.C. Beers\inst{2}         
\and C. Thom\inst{3}            
\and R. Wilhelm\inst{4}         
\and S. Rossi\inst{5}           
\and C. Flynn\inst{6}           
\and L. Wisotzki\inst{7}        
\and D. Reimers\inst{1}         
}

\offprints{N. Christlieb,\\ \email{nchristlieb@hs.uni-hamburg.de}}

\institute{
       Hamburger Sternwarte, Universit\"at Hamburg, Gojenbergsweg 112,
       D-21029 Hamburg, Germany;\\
       \email{nchristlieb/dreimers@hs.uni-hamburg.de} 
  \and Department of Physics \& Astronomy and JINA: Joint Institute for
       Nuclear Astrophysics, Michigan State University, East Lansing, MI
       48824, U.S.A.; \email{beers@pa.msu.edu} 
  \and Centre for Astrophysics and Supercomputing, Swinburne University of
       Technology, Hawthorn, VIC 3122, Australia; 
       \email{cthom@astro.swin.edu.au} 
  \and Department of Physics, Texas Tech University, Lubbock, TX 79409, U.S.A;
       \email{ron.wilhelm@ttu.edu} 
  \and Departamento de Astronomia Instituto de Astronomia, Geof{\'\i}sica e 
       Ci{\^e}ncias Atmosf{\'e}ricas, Universidade de S{\~a}o Paulo, 05508-900 
       S{\~a}o Paulo SP, Brazil; \email{rossi@astro.iag.usp.br} 
  \and Tuorla Observatory, University of Turku, V\"ais\"al\"antie 20, 21500 
       Piikki\"o, Finland; \email{cflynn@utu.fi} 
  \and Astrophysikalisches Institut Potsdam, An der Sternwarte 16, D-14482 
       Potsdam, Germany; \email{lutz@aip.de} 
}

\date{Received 11 August 2004 / Accepted 20 October 2004}

\abstract{ 
  We present a sample of 8321 candidate Field Horizontal-Branch (FHB) stars
  selected by automatic spectral classification in the digital data base of
  the Hamburg/ESO objective-prism survey. The stars are distributed over 8225
  square degrees of the southern sky, at $|b| \gtrsim 30\,\deg$. The average
  distance of the sample, assuming that they are all FHB stars, is $9.8$\,kpc,
  and distances of up to $\sim 30$\,kpc are reached.  Moderate-resolution
  spectroscopic follow-up observations and $UBV$ photometry of 125 test sample
  stars demonstrate that the contamination of the full candidate sample with
  main-sequence A-type stars is $< 16$\,\%, while it would be up to 50\,\% in
  a flux-limited sample at high galactic latitudes. Hence more than $\sim
  6800$ of our FHB candidates are expected to be genuine FHB stars.  The
  candidates are being used as distance probes for high-velocity clouds and
  for studies of the structure and kinematics of the Galactic halo.
}

\titlerunning{HES FHB stars}
\authorrunning{Christlieb et al.}
\maketitle

\section{Introduction}

Over the past few decades, as knowledge of the structure and nature of the
stellar components of the Milky Way has increased, the level of complexity
that has been revealed has also risen. Not only must astronomers attempt to
describe and quantify the global properties of our Galaxy, but they also must
contend with the presence (likely only partially recognized at present) of
real structures -- coherent streams arising from shredded dwarf galaxies, such
as Sagittarius \citep[e.g.,][]{Ibataetal:1994,Majewskietal:2003}, and the
still poorly-constrained Monoceros stream
\citep{Newbergetal:2002,Yannyetal:2003,Ibataetal:2003,Frinchaboyetal:2004,Martinetal:2004},
as well as the tidal tails of individual globular clusters \citep[][and
references therein]{Dehnenetal:2004}. Even within recognized stellar
components, such as the thick disk, evidence exists for the presence of
multiple populations, e.g., the so-called ``metal-weak thick disk''
\citep[see][and references therein]{Beersetal:2002} perhaps arising from stars
donated to the original thick disk by ravaged dwarf satellites. The overall
shape of the inner and outer halo of the Galaxy are still not well known,
though most recent evidence suggests that the outer halo is at least somewhat
flattened \citep[e.g.,][]{Chiba/Beers:2000,Helmietal:2003}.

It is clear that, for progress to be made, further detailed examination of the
positions, motions, and chemical compositions of the stars of the Milky Way is
required. Although not ideal for examination of elemental abundances (due to
their higher temperatures, and hence weaker metallic lines), Field
Horizontal-Branch (hereafter, FHB) stars can provide valuable tracers of the
structure, kinematics, and dynamics of the halo and thick disk of the
Galaxy. They are numerous throughout the Galaxy, and sufficiently luminous to
be identified up to large distances from the Sun. 

One immediate application of a large sample of tracers located throughout the
halo of the Galaxy is the opportunity to obtain dynamical estimates of its
total mass. \citet{Wilkinson/Evans:1999} used the full set of 27 known
satellite galaxies and globular clusters located at distances greater than
20\,kpc to explore this question. For six of these objects proper motions are
known. This data was later supplemented by \citet{Sakamotoetal:2003}, who
added some 400 FHB stars (at distances less than 10\,kpc) with available
radial velocities, at least 50\,\% of which also had available proper motions,
in order to refine the mass estimate of the Galaxy. This sample was further
expanded to include FHB stars from the early data release of the Sloan Digital
Sky Survey \citep[SDSS;][]{Yorketal:2000}, adding stars up to 75\,kpc away
\citep{Beersetal:2003}. Most recently, \citet{Clewleyetal:2004} have
identified a new sample of 60 FHB stars located in six high Galactic latitude
fields up to 50\,kpc away, which will be combined with additional distant
samples to obtain an improved estimate of the mass of the Galaxy.

In addition, FHB stars are of great importance as probes of the distances to
High Velocity Clouds (HVCs). HVCs are clouds of neutral hydrogen at velocities
incompatible with Galactic differential rotation \citep[see][for a recent
review]{Wakker:2004}. With HVCs, we might observe a continuous infall of
metal-poor ($\sim 1/10$ solar) gas into the Galaxy, which dilutes the
interstellar medium (ISM) \citep{Wakkeretal:1999}. A clearer understanding
of the nature of HVCs would therefore be important for modeling the chemical
evolution of the Galaxy. 

Distances to HVCs can be determined by using stars of known distance in the
line of sight to the clouds \citep{Wakker/vanWoerden:1997}. Provided that the
HVC under consideration has a detectable metal content, one can observe
absorption lines of these metals at the velocity of the cloud in the spectra
of stars located behind the cloud, but these same lines would not be seen in
the spectra of stars located in front of the cloud. Therefore, one can
constrain the HVC distance by using a sample of stars with different distances
along the line of sight to the cloud. FHB stars are particularly well-suited
for this purpose, because they are numerous, distant, and their spectra have
only very weak intrinsic absorption lines of metals.

Previously, the largest catalogues of candidate FHB stars have come from
objects selected on the basis of their distinctive appearance on
objective-prism plates (e.g., the HK survey of Beers and collaborators;
\citealt{Beersetal:1996}). The HK survey possesses a magnitude limit, defined
here as the $B$ magnitude of the faintest objects present in the survey, of
$B\simeq 15.5$, with a corresponding distance limit for FHB stars on the order
of 10\,kpc from the Sun. In the Hamburg/ESO survey \citep[HES;
][]{hespaperI,hespaperIII}, the fainter limiting magnitude of $B\simeq 18.0$
enables detection of FHB stars at distances of up to $\sim 30$\,kpc.

\begin{figure}[htbp]
  \begin{center}
    \epsfig{file=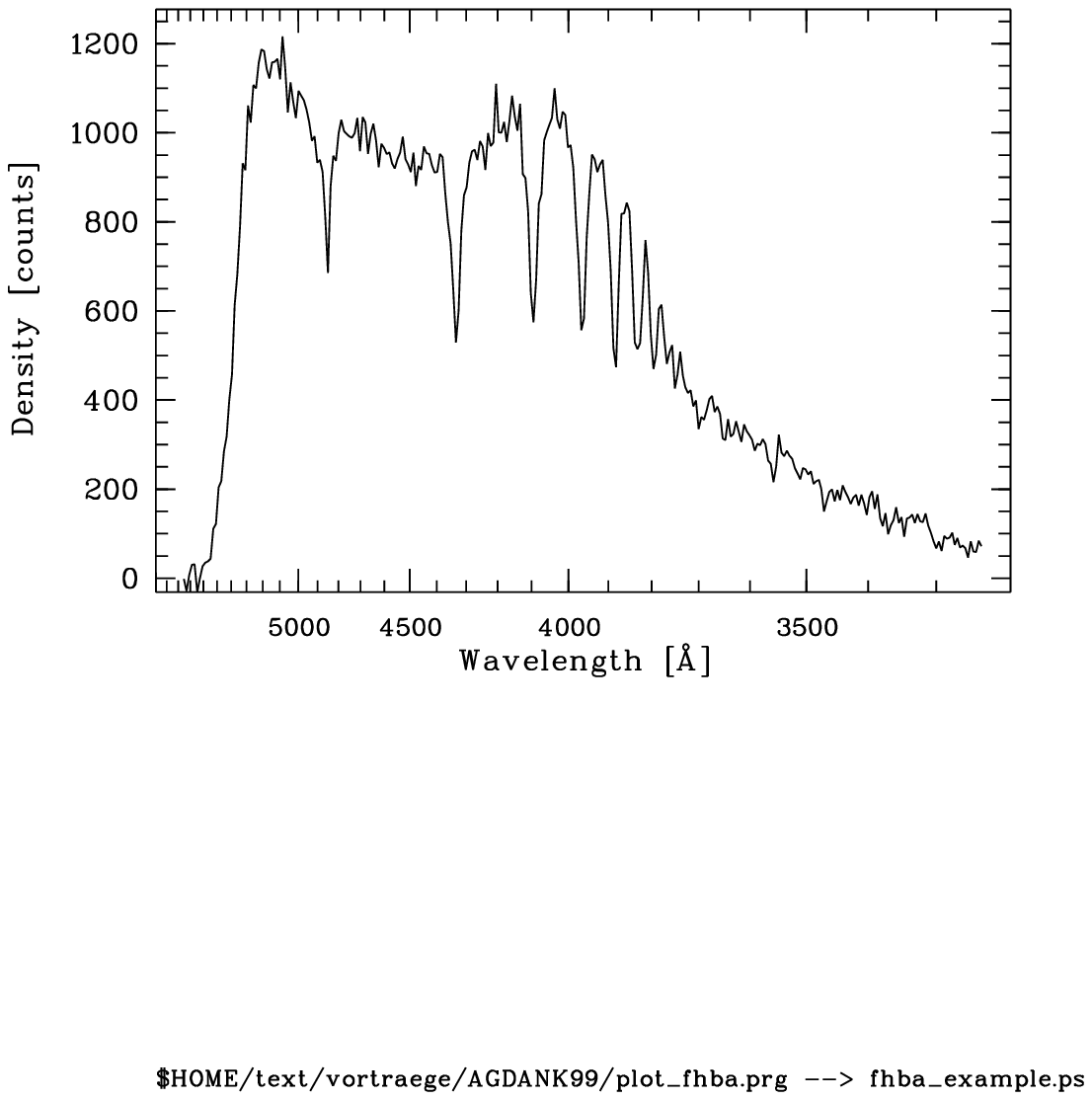, clip=, width=8.8cm,
      bbllx=125, bblly=441, bburx=421, bbury=641}
    \caption{\label{Fig:fhba_example} Sample HES prism spectrum of a
             candidate FHB star. Note that wavelength increases from
             right to left. The sharp drop at about 5400\,{\AA} is due to the
             sensitivity cutoff of the IIIa-J emulsion. }
  \end{center}
\end{figure}

Colorimetric surveys, such as the SDSS, are now able to identify FHB
candidates up $\simeq 100$\,kpc away
\citep{Yannyetal:2000,Sirkoetal:2004}. The 2MASS all-sky survey also provides
a valuable resource for the colorimetric identification of FHB
candidates. \citet{Brownetal:2004}, for example, show that 2MASS $JHK$
photometry can be used to efficiently select FHB candidates out to 10\,kpc
from the Sun. These authors provide a catalogue of some 100,000 FHB candidates
located above Galactic latitude $|b| = 15\deg$ (note, however, the
contamination issues described below). Many more are identifiable at lower
latitudes; however, the patchiness and increased overall level of interstellar
reddening presents additional difficulties. The 2MASS catalogues of FHB
candidates provide a valuable complement to the SDSS approach, since the
bright limit of SDSS ($B \simeq 14.5$) is similar to the faint limit of the
2MASS FHB candidates. In due course, the all-sky survey presently underway in
the far-UV and near-UV with the GALEX mission will provide another valuable
source of FHB candidates.

It must be kept in mind that catalogues of FHB candidates can be easily
confounded with high surface gravity A-type stars. A number of authors have
pointed out that roughly 50\,\% of all stars with colors that fall in the same
range as FHB stars are in fact of this lower luminosity variety, many of which
are likely to be halo blue stragglers \citep[see,
e.g.,][]{Norris/Hawkins:1991,Prestonetal:1991,Prestonetal:1994,Wilhelmetal:1999b}.
The majority of the so-called ``blue metal poor'' stars of
\citet{Prestonetal:1994} are also likely to be blue stragglers
\citep{Preston/Sneden:2000}. High-gravity A-type stars can be distinguished
from lower gravity FHB stars through the use of broadband $UBV$ photometry
combined with medium-resolution spectroscopy \citep{Wilhelmetal:1999a}, due to
the large effect of surface gravity on the broadband $U$ flux, and on the
wings of the Balmer lines. \citet{Clewleyetal:2002} describe procedures that
attempt to make this separation based on medium-resolution spectroscopy alone.

In the HES we have attempted to select a \emph{clean} sample of FHB
candidates; that is, we wanted to keep the ``contamination'' with
main-sequence stars as low as possible. If we can demonstrate by
representative follow-up observations that the contamination is low, such a
selection would allow us to isolate a large, statistical sample of FHBs,
useful for many purposes, directly from the HES data base, without the need of
any follow-up observations.

\begin{figure*}[htbp]
  \leavevmode
  \begin{center}
    \epsfig{file=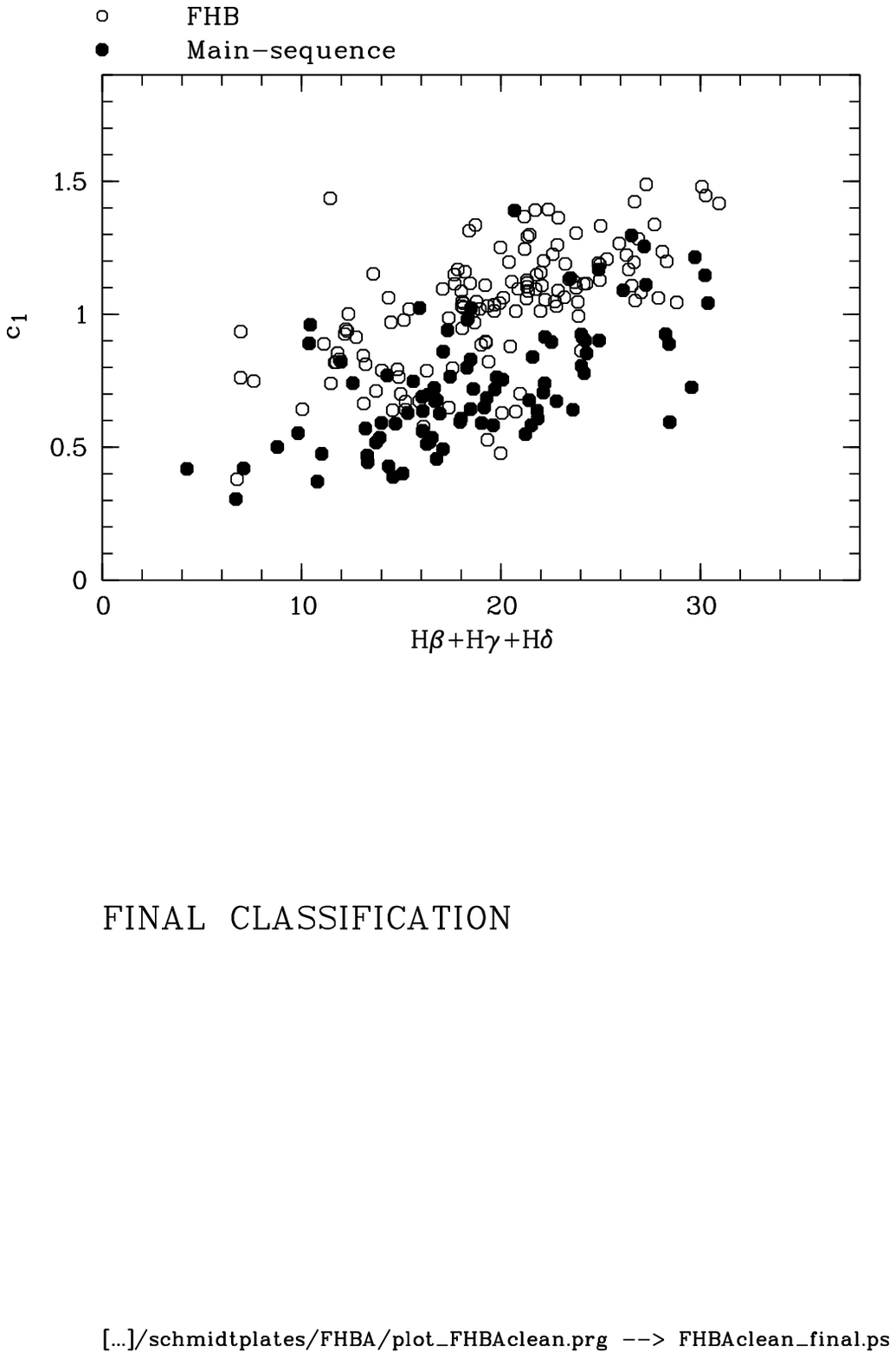, clip=, width=8.8cm, bbllx=107, bblly=513,
      bburx=399, bbury=757}\hspace{0.3cm}
    \epsfig{file=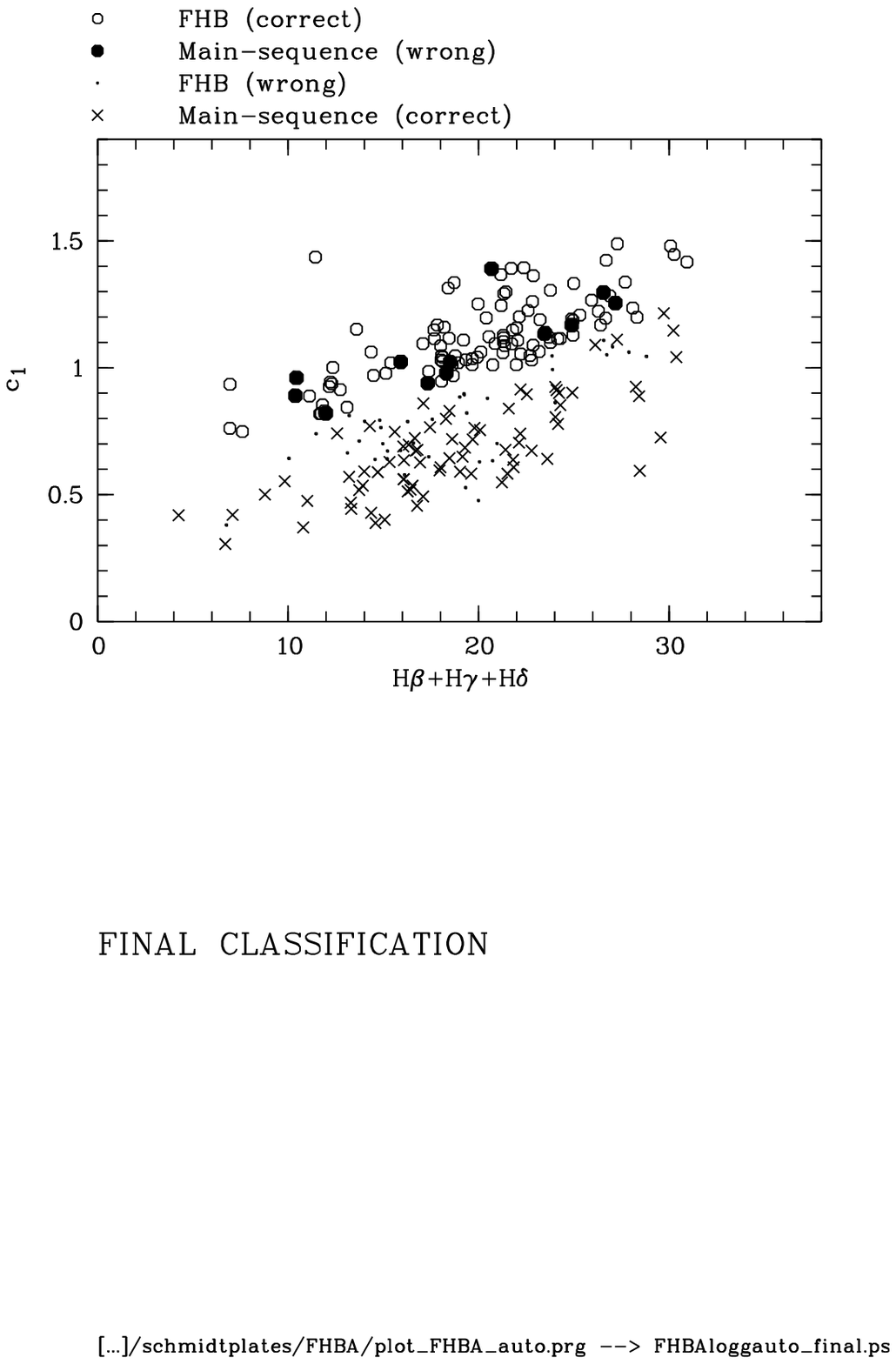, clip=, width=8.8cm, bbllx=107, bblly=513,
      bburx=399, bbury=757}
    \caption{\label{Fig:FHBAclassification} Automatic classification of
       A-type stars in the two-dimensional feature space $c_1$ versus
       sum of Balmer line equivalent widths
       $\mbox{H}\beta+\mbox{H}\gamma+\mbox{H}\delta$. In the left panel we
       show the learning sample of 228 stars before the classification is
       applied; in the right panel, crosses and dots denote the main-sequence
       and FHB stars that have been classified as main-sequence stars,
       respectively, and empty and full circles are the objects that have been
       classified as FHB stars.}
  \end{center}
\end{figure*}

In Section \ref{Sect:FHBAsel} we describe our procedure for selecting FHB
candidates. In Section \ref{Sect:Results}, the results of the application of
the selection algorithm to a major fraction of the HES data base (i.e.,
spectra from 329 out of a total of 380 HES fields) are presented. We also
demonstrate that a low contamination with high-gravity A-stars can indeed be
achieved with this selection. We present our conclusions in Section
\ref{Sect:Conclusions}.

\section{Selection of FHB candidates}\label{Sect:FHBAsel}

A-type stars are readily detectable in the HES due to their strong Balmer
lines (for an example see Figure \ref{Fig:fhba_example}). FHB candidates are
selected in the HES in two steps. First, we apply $B-V$ and $U-B$ cutoffs, and
then we employ automatic spectral classification to seperate main-sequence
from horizontal-branch stars. The classification is based on the Str\"omgren
$c_1$ colour index and the sum of the equivalent widths of the Balmer lines
H$\beta$, H$\gamma$ and H$\delta$. All of this information is derived directly
from the HES spectra. The calibrations of $B-V$, $U-B$ and $c_1$ are accurate
to $0.1$, $0.09$ and $0.15$\,mag, respectively \citep{HESStarsI}. The Balmer
line strengths are measured with an iterative algorithm in which a Gaussian
profile is fitted to the lines \citep{HESStarsI}. We note that using a
Gaussian profile may lead to systematic underestimation of the strengths of
the broad lines of hot stars. However, for our FHB candidate selection, the
internal consistency of the measured values rather than their absolute scale
is important. We estimate the internal accuracy of the equivalent width sum
$\mbox{H}\beta+\mbox{H}\gamma+\mbox{H}\delta$, as measured from the HES
spectra, to be on the order of a few {\AA}ngstr\"oms.

In the first step of the procedure, we select all stars that lie in the colour
range
\begin{equation}
-0.2 < B-V < 0.3\label{Eq:BVcrit}
\end{equation}
and
\begin{equation}
-0.3 < U-B < 0.5\label{Eq:UBcrit}.
\end{equation}
The $B-V$ criterion selects all stars with effective temperatures in the range
$7000\,\mbox{K} \lesssim \teffm \lesssim 10,000\,\mbox{K}$, while allowing for
an error margin of 0.1\,mag in $B-V$.  The $U-B$ criterion was chosen to
include most FHB stars, while a potential contamination with DA white dwarfs
is kept at a minimum (see Figure 13 of \citealt{HESStarsI}).

In order to develop a method for the separation of main-sequence from
horizontal-branch stars, we investigated a set of 259 HES spectra of stars
that were analysed with the methods of \citet{Wilhelmetal:1999a}. This sample
includes 214 stars from \citet{Wilhelmetal:1999b} present on the HES
plates. The remaining 45 stars resulted from our own follow-up spectroscopy
and photometry of candidates selected with a preliminary selection
algorithm. From this set of stars, 31 were classified either as ``Am'' (i.e.,
metallic-line A-type stars), ``Ap'' (i.e., peculiar A-type stars), or
``FHB/A'' (i.e., no definitive conclusion could be reached as to whether these
stars are on the horizontal-branch or are main-sequence stars). These were
discarded, leaving 90 high-gravity A-type stars and 138 FHB stars.

Inspection of the distribution of these stars in a $c_1$ versus
$\mbox{H}\beta+\mbox{H}\gamma+\mbox{H}\delta$ diagram reveals that these two
classes of objects are reasonably well separated in this feature space (see
Figure \ref{Fig:FHBAclassification}). We hence implemented an automatic
spectral classification algorithm that makes use of these two features.  The
classification forms the second step of our FHB candidate selection, i.e., it
is applied only to the subset of the objects in the HES data base that meet
criteria (\ref{Eq:BVcrit}) and (\ref{Eq:UBcrit}).

A detailed description of the classification algorithm can be found in
\cite{HESStarsII}.  The above described sample of 90 confirmed high-gravity
A-type stars and 138 FHB stars is used as a training set. Based on this
sample, the probability distributions for objects being member of one of the
two classes, given their $c_1$ and Balmer line equivalent width sums, are
estimated. These so-called class-conditional probability distributions are
modeled by two-dimensional Gaussian distributions. This results in
iso-probability lines in the two-dimensional feature space which have the
shapes of ellipses, with their center being located at the point with the
coordinates of the class means of $c_1$ and
$\mbox{H}\beta+\mbox{H}\gamma+\mbox{H}\delta$. For example, the mean of $c_1$
for the main-sequence stars in the left panel of Figure
\ref{Fig:FHBAclassification} is $0.7$\,mag, and for the Balmer line sum it is
$19$\,{\AA}.

\begin{figure}[htbp]
  \leavevmode
  \begin{center}
    \epsfig{file=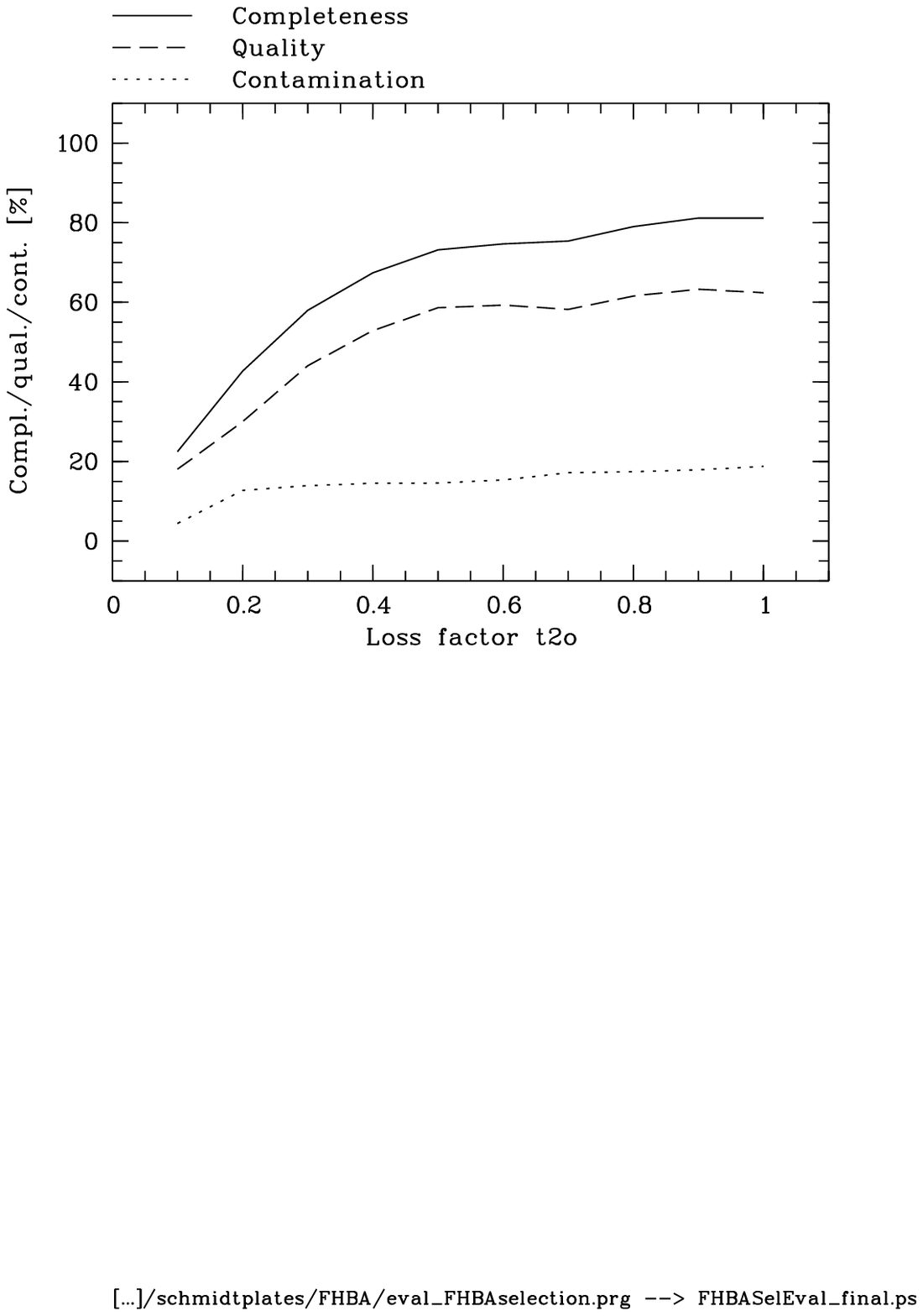, clip=, width=8.8cm, bbllx=73, bblly=372,
      bburx=372, bbury=604}
    \caption{\label{Fig:costfac_adjust} Adjusting the cost factor \texttt{t2o}
      by evaluating the classifcation results. For further explanation see text.}
  \end{center}
\end{figure}

Our statistical classification requires an estimation of the prior
probabilities of the two classes considered here, i.e., main-sequence or FHB
star.  \citet{Wilhelmetal:1999b} analysed a sample of 1121 A-type stars
selected from HK survey plates. We consider it to be a representative
flux-limited sample at high galactic latitude, and use it to estimate the
prior probabilities of ``FHB'' and ``A'' stars in the
HES. \citet{Wilhelmetal:1999b} find 416 ``FHB'' and 444 ``A'' stars in their
sample. Therefore, we estimate the prior probabilities for these stars to be
0.396 and 0.371 in the HES, respectively.

Any given object with measured values for $c_1$ and
$\mbox{H}\beta+\mbox{H}\gamma+\mbox{H}\delta$ can then be assigned to one of
the two classes with various decision rule. If the so-called Bayes' rule is
used, an object is assigned to the class with the highest posterior
probability, given its values of the two features. This decision rule results
in ``dividing lines'' in the two-dimensional feature space that have the shape
of a polynomial of second order (hence this classification method is called
``quadratic discriminant analysis''). From looking at the right panel of
Figure \ref{Fig:FHBAclassification}, one can have a presentiment of the
parabola-shaped selection boundary for our classification problem.

Classification with Bayes' rule minimizes the total number of
misclassifications, if the true distributions of the class-conditional
probabilities are used. However, in our case our classification aim is to
compile a clean sample of FHB stars, which is different from minimizing the
total number of misclassifications, because different classification errors
will have different weight. For example, a misclassification of a high-gravity
A-type star as an FHB star is worse for the compilation of a clean sample than
a misclassification in the other direction.

In order to to reach our aims, we have employed a so-called minimum cost rule
classification, which minimizes the total cost. Here, every misclassified
object causes a cost, while a correctly classified object is cost-neutral.
The choice of cost factors determines the relative weight of
misclassifications from one class to another. As described in
\citet{HESStarsII}, the cost factor \texttt{t2o} determines the cost for
misclassifications of an object of the {\bf t}arget class (`\texttt{t}') to
(`\texttt{2}') one of the {\bf o}ther classes (`\texttt{o}'). With the
cost factor \texttt{o2t}, the cost for the contamination of the target class
is adjusted, and \texttt{o2o} is the cost factor for misclassifications
between other classes. Since we are here only dealing with two classes -- that
is, the target class of FHB stars, and high-gravity A-type stars -- the cost
factor \texttt{o2o} does not have any meaning. Therefore, effectively only one
cost factor has to be chosen. As explained in \cite{HESStarsII}, only the
\emph{relative} cost factor values need to be adjusted. Therefore, we set
\texttt{o2t} to a fixed value of $1.0$ and investigated the classification
results as a function of \texttt{t2o}, with $0.0 \leq \mbox{\texttt{t2o}} \leq
1.0$. The result is shown in Figure \ref{Fig:costfac_adjust}.

\begin{figure*}[htbp]
  \leavevmode
  \begin{center}
    \epsfig{file=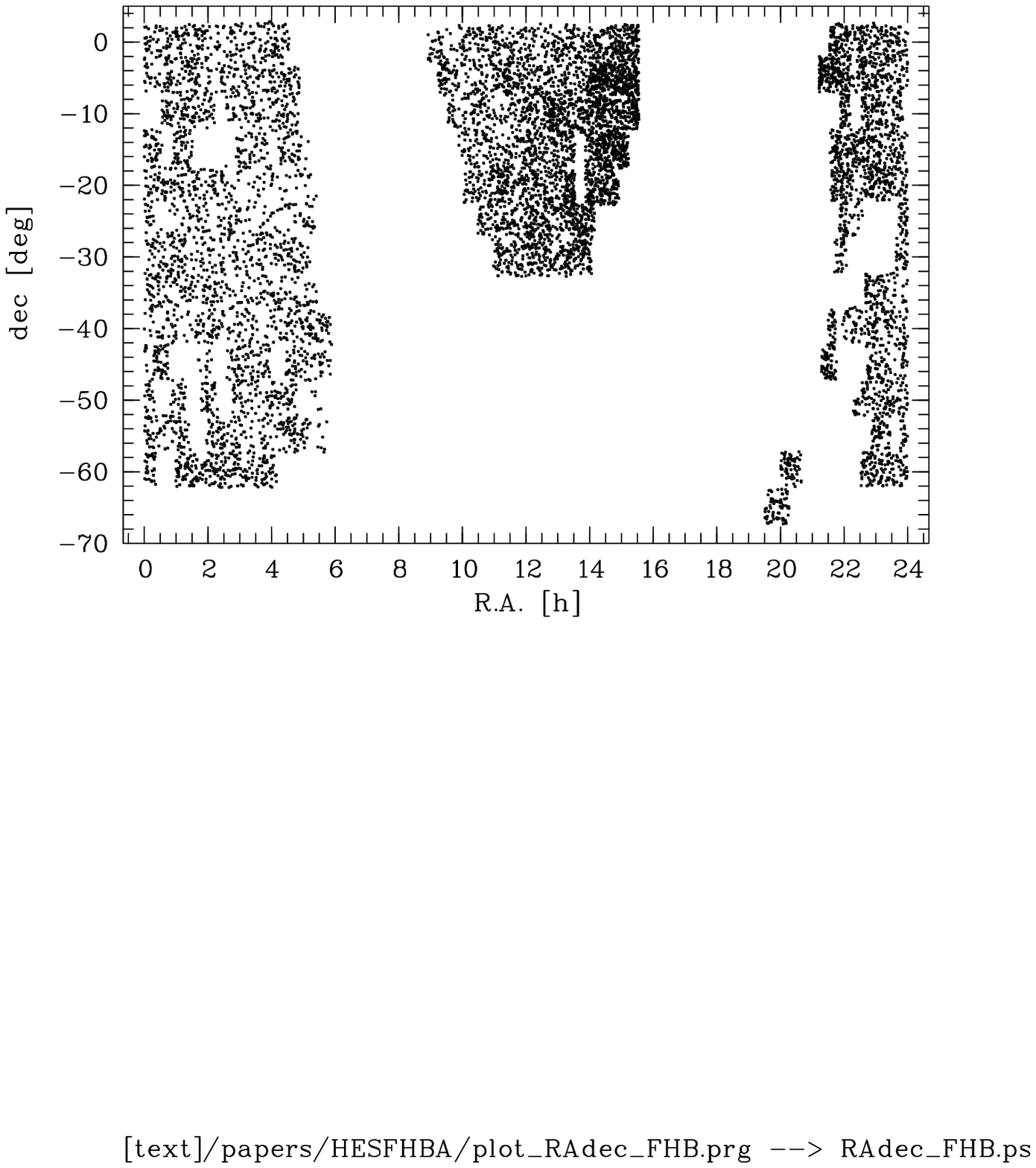, clip=, width=8.8cm, bbllx=89, bblly=335,
      bburx=484, bbury=598}\hspace{0.3cm}
    \epsfig{file=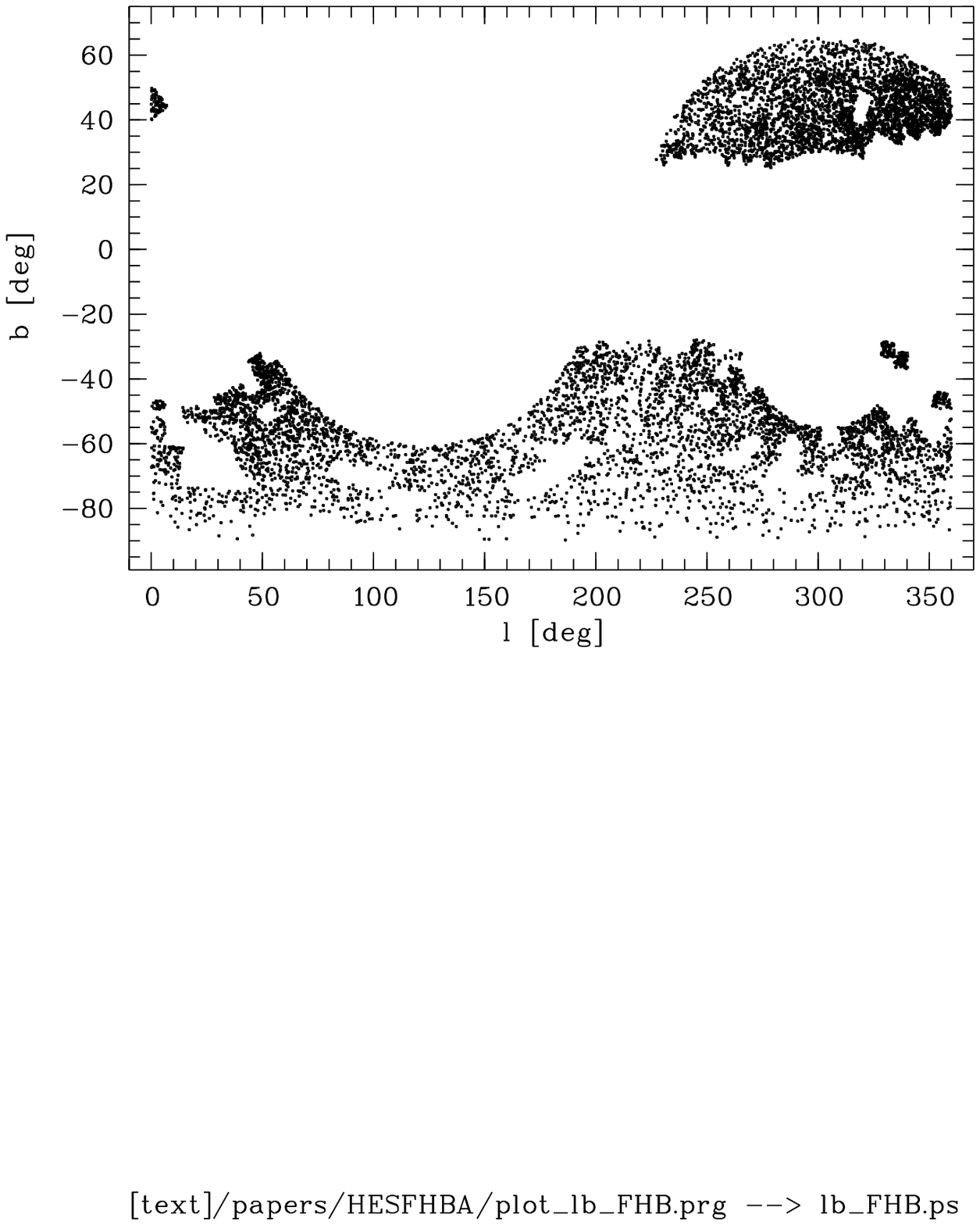, clip=, width=8.8cm, bbllx=89, bblly=335,
      bburx=484, bbury=598}
    \caption{\label{Fig:FHB_sky_distrib} Distribution of the 8321 HES FHB
    candidates on the sky.}
  \end{center}
\end{figure*}

We define the quality $q$ of the classification as the completeness rate minus
the contamination rate. This gives $q = 100$\,\% for the perfect
classification with respect to our classification aim, i.e., 100\,\%
completeness (all FHB stars classified correctly) and 0\,\% contamination (no
high-gravity A-type star erroneously classified as an FHB star). The highest
value of $q$ which is reachable is a measure of how well the objects of the
target class can be seperated from objects of the other class(es).

In our case, the highest quality is achieved at $\mbox{\texttt{t2o}}\sim
0.9$. However, we decided to adopt $\mbox{\texttt{t2o}}=0.5$ for our
classification, in order to keep the high-gravity A-type star contamination
below 15\,\%. Estimates of the expected error rates with the leaving-one-out
method \citep[for a description see][]{HESStarsI} for the adopted value of
\texttt{t2o} yield an expected completeness of the FHB sample of 73\,\%.

\section{Results}\label{Sect:Results}

\subsection{Application to 329 HES fields}

Application of our selection algorithm to the 329 HES fields that are
currently used for the exploitation of the stellar content of the HES yielded
8321 FHB candidates. As we argue in Section \ref{Sect:Contamination} below,
statistically more than $\sim 6800$ of these are expected to be genuine FHB
stars. The distribution of these objects on the sky is shown in Figure
\ref{Fig:FHB_sky_distrib}. In Table A.1 we list, for the entire sample,
photometry ($UBV$, $c_1$), and Balmer line equivalent width sums, as derived
directly from HES spectra, as well as coordinates of the stars.

\begin{figure}[htbp]
  \leavevmode
  \begin{center}
    \epsfig{file=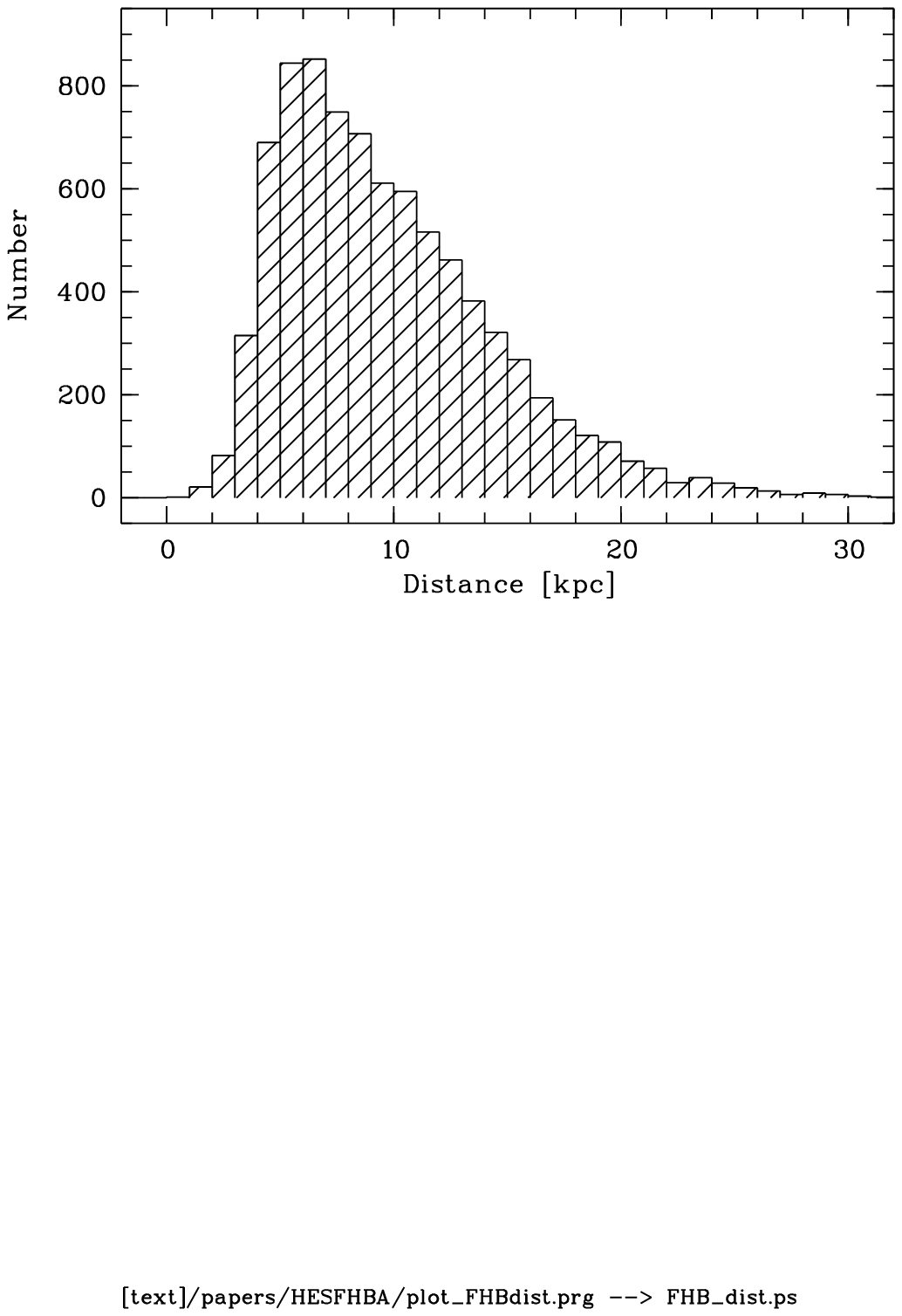, clip=, width=8.8cm, bbllx=68, bblly=427,
      bburx=371, bbury=626}
    \caption{\label{Fig:FHB_dist} Distance distribution of the FHB candidate
    sample of 8321 stars, assuming that they are all indeed FHB stars.}
  \end{center}
\end{figure}

We determined the distance distribution of the sample by assuming that all
candidates are indeed FHB stars. Strictly speaking, this assumption is not
valid, because our candidate sample will contain some non-FHB stars. However,
this would change the distance distribution noticably only if the apparent
magnitude distribution of the non-FHB contaminants were significantly
different from that of the FHB stars (for which we do not have any indication
from our sample of non-FHB stars in our training set), and if the fraction of
contaminants were high (we will demonstrate the opposite below).  The
reddening of the stars was deduced from the maps of \citet{Schlegeletal:1998},
using the empirical formula of \citet{Beersetal:2002} to correct for the
suspected overestimate of the reddening by Schlegel et al. in the cases were
$E(B-V) > 0.1$. This correction had to be applied to only 6\,\% of our stars;
the average reddening of our sample is only $E(B-V) = 0.046$. Therefore, the
extinction corrections are negligible (i.e., the distances would change on
average by $\sim 2$\,\%) compared to other sources of error in our rough
distance estimates.

Estimates of the absolute visual magnitudes, $M_V$, of our FHB candidates are
obtained following the procedures of \citet{Prestonetal:1991}. These authors
assembled a mean $M_V$ versus $B-V$ relation for blue horizontal branch stars
in 15 Galactic globular clusters, normalized relative to an assumed absolute
magnitude of RR Lyrae stars in each cluster. The metallicity dependence is
accounted for using the corrections of \citet{Clementinietal:1999} for RR
Lyrae stars. Since [Fe/H] is not available for the full sample of HES FHB
candidates, we assume that they all have the average value of
\citet{Wilhelmetal:1999b}, i.e., $\mbox{[Fe/H]}=-1.9$. The correction is
0.36\,mag with respect to assuming solar metallicity (that is, the absolute
$V$ magnitudes of the metal-poor FHB stars are brighter by this amount),
resulting in distances increased by 18\,\%. We then use the de-reddened HES
$B-V$ colours and $V$ magnitudes to compute distance moduli, taking into
account extinction.

The average distance of our sample, assuming that all of the candidates are
FHB stars, is 9.8\,kpc, while distances of up to 30\,kpc are reached (see
Figure \ref{Fig:FHB_dist}).

\subsection{Contamination with high-gravity A-type stars}\label{Sect:Contamination}

For 125 stars of our sample that are not part of the learning sample we have
taken moderate-resolution ($\sim 2$\,{\AA}) spectra with the CTIO 4\,m/RCS in
December 1999, and the ESO 3.6\,m/EFOSC2 in November 2000. The spectra have
signal-to-noise ratios of $S/N > 20$ per pixel at \ion{Ca}{ii}~K. After the
standard CCD reductions, the algorithm of \citet{Horne:1986} was employed to
extract the spectra in order to maximize their $S/N$. $UBV$ photometry was
obtained at the ESO-Danish 1.54\,m with DFOSC in observing runs in November
1998 and November 2000. The results of the photometry runs will be reported in
Beers et al. (in preparation).

The spectra and photometry was analysed with the method of
\citet{Wilhelmetal:1999a}, in which the observed Balmer line profiles,
equivalent widths of \ion{Ca}{ii}~K, and $UBV$ photometry are compared to the
respective properties of synthetic spectra. This technique allows the
estimations of effective temperature {\tefft}, surface gravity $\log g$, and
metallicity [Fe/H] with accuracies of $\sigma\left(\teffm\right)=\pm 250$\,K,
$\sigma\left(\log g\right)=\pm 0.14$\,dex, and
$\sigma\left(\mbox{[Fe/H]}\right)=\pm 0.12$\,dex, which permits one to
reliably seperate main-sequence A-type stars from FHB stars.

\begin{table}
  \centering
  \caption{\label{Tab:TestSampleEvaluation} Evaluation of the test sample of
  125 FHB candidates.}
  \begin{tabular}{lrr}\hline
    Class & \multicolumn{1}{l}{$N$} & \multicolumn{1}{l}{Fraction}
    \rule{0.0ex}{2.3ex}\\\hline
    FHB   &  103 & 82\,\% \rule{0.0ex}{2.3ex}\\
    FHB/A &    5 &  4\,\% \\
    A     &   15 & 12\,\% \\
    Am    &    2 &  2\,\% \\\hline
  \end{tabular}
\end{table}

The result is shown in Table \ref{Tab:TestSampleEvaluation}. A total of 103
stars (or 82\,\%) were confirmed to be FHB stars, and a further 5 stars were
classified as ``FHB/A'' stars. Since the ``FHB/A'' stars could be either
``FHB'' or ``A'' stars, the total fraction of FHB stars lies in the range
82--86\,\%. A total of 15 stars (or 12\,\%) were classified ``A'' by the
method of \citet{Wilhelmetal:1999a}, yielding a total contamination of
12--16\,\%, which agrees very well with the contamination of $\sim 15$\,\%
expected from the training of the classifier applied to the HES data base.

\section{Conclusions}\label{Sect:Conclusions}

This large sample of HES FHB candidates, homogeneously distributed over the
high-Galactic latitude sky, is of great value for studies of the structure and
kinematics of the Galactic halo. The stars are also being used as distance
probes for HVCs. In a forthcoming paper we will report on a cross-correlation
of the sample with HVC moment maps generated from data of the H~I Parkes
All-Sky Survey (Thom et al., in preparation).

This sample of HES FHB candidates is expected to contain a contamination with
main-sequence A-type stars of less than 16\,\%, so that for many studies the
stars can be used directly, without the need for extensive follow-up
observations to further refine the classifications.

\begin{acknowledgements}
We thank A. Frebel for technical help and careful proof-reading.
T.C.B. acknowledges partial support from grants AST 00-98508 and AST 00-98548,
and PHY 02-16783, Physics Frontier Centers/JINA: Joint Institute for Nuclear
Astrophysics, awarded by the US National Science Foundation. C.T. acknowledges
the financial support of the Australian Research Council through its Linkage
International program. S.R. thanks the Brazilian Institutions FAPESP and CNPq
for partial financial support. C.F.  thanks the Academy of Finland for its
support through the Antares program for space research.
    
\end{acknowledgements}

\bibliography{classification,datanaly,datared,fhba,HES,mphs,ncpublications,ncastro,statistics,wd}
\bibliographystyle{aa}

\begin{appendix}

  \section{The HES FHB candidates}

  In Table A.1 we list details of the sample of 8321 HES FHB candidates stars
  described in this paper. The table is made available only electronically. It
  contains the following columns:

  \begin{flushleft}
  \begin{tabular}{ll}
    hename   & HE designation\\
    ra2000   & R.A. at equinox 2000.0, derived from DSS~I\\
    dec2000  & Declination at equinox 2000.0, derived from DSS~I\\
    field    & ESO-SERC field number\\
    $B$      & $B$ magnitude\\ 
    $B-V$    & $B-V$ magnitude, measured from HES spectra\\
    $U-B$    & $U-B$ magnitude, measured from HES spectra\\
    $c_1$    & Str\"omgren $c_1$ index, measured from HES spectra\\
    balmsum  & Sum of equivalent widths of $\mbox{H}\beta$, $\mbox{H}\gamma$,
               and $\mbox{H}\delta$\\
  \end{tabular}
  \end{flushleft}
  
\end{appendix}

\end{document}